\documentclass[a4paper,11pt]{article}
\usepackage{pos}
\usepackage{amsmath}
\usepackage{graphicx}
\usepackage{hyperref}
\usepackage{multicol}

\title{NLO Effects in QCD Sum-Rule Analyses of $f_{0}(500)$ as a Tetraquark state}

\author*[a,b]{B.A. Cid-Mora}
\author[a]{T.G. Steele}

\affiliation[a]{Department of Physics \&
Engineering Physics, University of Saskatchewan, SK, S7N 5E2, Canada}
\affiliation[b]{GSI Helmholtzzentrum f\"{u}r Schwerionenforschung, Darmstadt, Germany}

\emailAdd{bac302@usask.ca}
\emailAdd{tom.steele@usask.ca}

\abstract{
QCD sum-rule studies have been useful to understand and get an insight on the structure of exotic states, such as tetraquark systems. Moreover, the majority of these studies are performed only at leading-order (LO) within the light tetraquarks systems picture, overlooking the effects of higher order corrections, thus motivating our analysis. Our study \cite{Cid-Mora:2022kgu} focused on the effects of next-to-leading order (NLO) contributions to the mass estimates of the lightest tetraquark state ($J^{PC} = 0^{++}$), the so-called $\sigma$ or $f_{0}(500)$\cite{ParticleDataGroup:2020ssz}, using ratios of QCD Laplace sum-rules. A variety of different models were used, which included multiple resonances and width effects, resulting in a final mass prediction of $0.52\,\text{GeV}< m_{\sigma}< 0.77\,\text{GeV}$. Even though the ratios of sum-rules demonstrated some insesitivity under superficially large NLO contributions, they added the beneficial feature of canceling the dependence on the anomalous dimension. Our findings were in good agreement with patterns found in Chiral Lagrangian studies regarding the four-quark structure of the $\sigma$ state, including the relative coupling strengths within the multiple resonance analysis.
}

\FullConference{%
The 39th International Symposium on Lattice Field Theory,\\
8th-13th August, 2022,\\
Rheinische Friedrich-Wilhelms-Universit\"{a}t Bonn, Bonn, Germany
}

\begin{document}
\maketitle
	
\section{Introduction}

The hadronic sector has benefited from a substantial progress since the development of collider physics, and the discovery of the exotic state $X(3872)$ \cite{Belle:2003nnu} in 2003 by Belle Collaboration, and all the recent observations such as $Z_{c}(4430)$, $Z_{c}(3900)$ \cite{Belle:2007hrb, BESIII:2013ris, Belle:2013yex}, and $T_{cc}^{+}$ \cite{LHCb:2021auc,LHCb:2021vvq} had opened the door to consider exotic structures that go beyond the conventional quark model \cite{Jaffe:1976ig} (\textit{e.g.}, tetraquarks). In the lightquark sector, the four-quark structure picture provides a compelling framework to study the inverted mass hierarchy \cite{Jaffe:1976ig} of the lightest scalar mesons.

QCD sum-rules (QCDSR) \cite{Shifman:1978bx,Shifman:1978by} have been used to study conventional states up to NLO, whereas the majority of exotic states had only been studied at LO. Therefore, given that four-quarks interpretation seem to be inevitable to explain the existence of certain states, and the need for deeper analysis of the light tetraquarks states at NLO, %and due to the low quantity of studies of light tetraquarks states at NLO
our paper \cite{Cid-Mora:2022kgu} focused on the lightest scalar $\sigma$ as a tetraquark state, and estimated the NLO effects on the mass prediction using QCDSR. The exact calculations of next-to-leading order perturbative terms (NLO PT) were calculated in \cite{Groote:2014pva}, while the leading-order (LO) terms, as well as an exhaustive analysis of the optimal currents to perform a QCDSR analysis, were made by \cite{Chen:2007xr}. Based on this results, we performed a detailed study of the effects of NLO contributions compared to the LO in QCDSR ratios, which led to mass predictions given a variety of resonance(s) models and width shapes.

\section{Methodolody Analysis QCD Laplace Sum-rule}

QCDSRs are a useful technique that, based on the concept of quark-hadron duality, aims to emphasize the low-lying states contained in the spectral function. The hadronic spectral function containing the states of interest, with physical threshold $t_{0}$, is connected to the correlation function via dispersion relation
\vspace{-0.3cm}
\begin{equation}
\Pi\left(Q^2\right)=\Pi(0)+Q^2\Pi'(0)+\frac{1}{2}Q^4\Pi''(0)
+\frac{1}{3}Q^6\Pi'''(0)
+Q^8
\int\limits_{t_0}^\infty
\,\frac{ \rho(t)}{t^4\left(t+Q^2\right)}dt\,.
\label{GenDispRel}
\end{equation}
Later, by applying the Borel transform operator to \eqref{GenDispRel}, we can get rid of unknown subtraction constants, resulting in a family of Laplace sum-rules, which involve the theoretical entity \cite{Shifman:1978bx,Shifman:1978by}
\vspace{-.3cm}
\begin{equation}
{\cal L}_{k}(\tau)=\int\limits_{t_0}^\infty
t^k e^{-t\tau}\rho(t)\,dt \,,~ k= 0,1,2,\ldots \, .
\label{GenLap}
\end{equation}
The final form of the Laplace sum-rules will be outlined in the next section. 

\subsection{NLO Effects: First Results}
The spectral hadronic function can be  expressed separately as perturbative (PT) and non-perturbative (non-PT) contributions, the latter also known as QCD condensates. For the states of interest, these condensate terms were calculated in \cite{Chen:2007xr}, and the NLO PT corrections in \cite{Groote:2014pva}, resulting in
\vspace{-.2cm}
\begin{gather}
\rho^{{\rm PT}}(s)= \frac{s^{4}}{11520\pi^6} + \frac{s^{4} }{11520\pi^6}
\frac{\alpha_{s}(\mu)}{\pi}\Biggl [\frac{409-192\sqrt{2}}{40} +\frac{7-6\sqrt{2}}{4}\log\Bigl(\frac{\mu^{2}}{s} \Bigr)\Biggr],
\label{rho_NLO}
\end{gather}
where only numerically significant quark mass terms are included.

In principle, the anomalous dimension term associated with \eqref{rho_NLO} prevents application of the established methodology for RG improvement of Laplace sum-rules for light-quark systems \cite{Narison:1981ts}. However, up to NLO the spectral function is modified as 
\vspace{-.3cm}
\begin{equation}
\tilde\rho(s)=    \left(\alpha_s\right)^{2\gamma_1/\beta_1} \rho(s)\,,\quad\mathrm{where}\quad \gamma_{1} = \frac{7-6\sqrt{2}}{4}\,.
\label{tilde_rho}
\end{equation}
\eqref{tilde_rho} implies that ratios of sum-rules can be safely used, since the term $\gamma_1/\beta_1$ cancels, thus motivating an analysis independent of the anomalous dimension focused on these sum-rule ratios. 

\begin{figure}[htb]
    \centering
    \includegraphics[scale=.8]{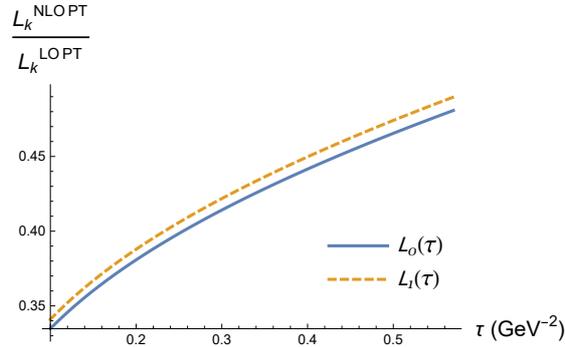}
    \caption{\small{The ratio of NLO and LO PT contributions to 
    ${\cal L}^{\rm QCD}_k(\tau)$ is shown as a function of $\tau$ for $k=0$ (solid blue curve) and $k=1$ (dashed orange curve).}} 
    \label{Lk_NLO_fig}
\end{figure}

Our first results showed large contribution of NLO terms to the Laplace sum-rules compared to LO alone in the spectral function (see Fig.~\ref{Lk_NLO_fig}), hence supporting the need for a comprehensive study of these NLO terms, and their impact on the mass prediction from the QCDSR approach.
\vspace{-.3cm}

\subsection{Borel Window: More Results}
The hadronic spectral function $\rho(t)$ with physical threshold $t_0$ % appropriate to the quantum numbers 
in \eqref{GenLap}, can also be split into a resonance contribution ($\rho^{\rm res}$) and a QCD continuum \cite{Shifman:1978bx,Shifman:1978by,Reinders:1984sr,Narison:2002woh,Gubler:2018ctz}, whose continuum threshold is denoted as $s_{0}$, hence resulting in the standard form of the Laplace sum-rule 
\vspace{-0.1cm}
\begin{equation}
{\cal R}_k\left(\tau,s_0\right)=\int\limits_{t_0}^{s_0}
t^k e^{-t\tau}\rho^{\rm res}(t)\,dt\,.
\label{LSR_final}
\end{equation}
The methodology is basically composed by the following steps:
First, we need to determine a reliable Borel window for the full analysis. 
Second, we must choose a variety of models that represent the resonance with their appropriate number of parameters; and finally we need a proper optimization method to obtain these parameters, and consequently get a mass prediction.

\begin{figure}[htb]
    \centering
    \begin{minipage}{.45\textwidth}
        \centering
        \includegraphics[scale=0.65]{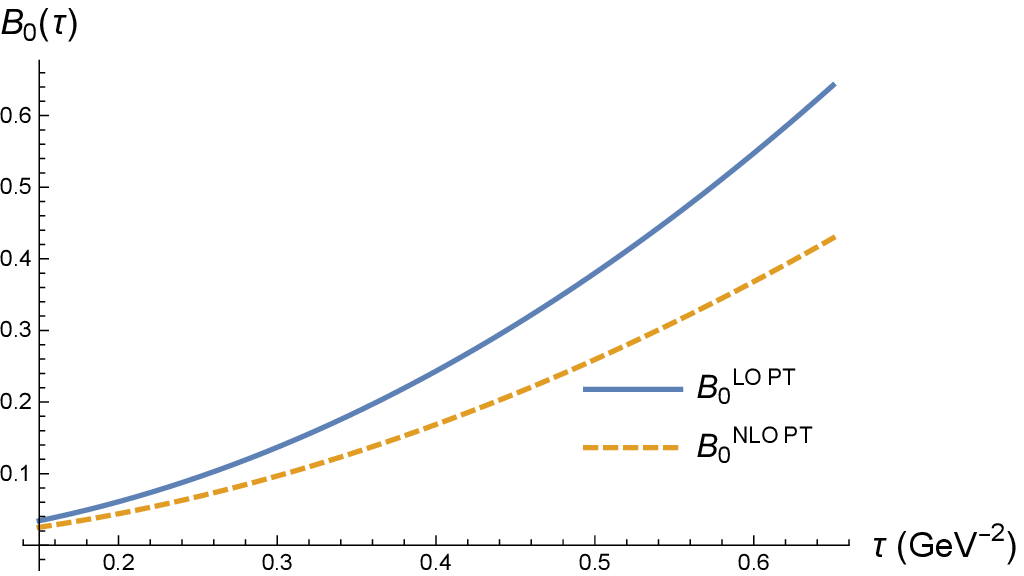}
    \end{minipage}
    \begin{minipage}{.45\textwidth}
        \centering
        \includegraphics[scale=0.65]{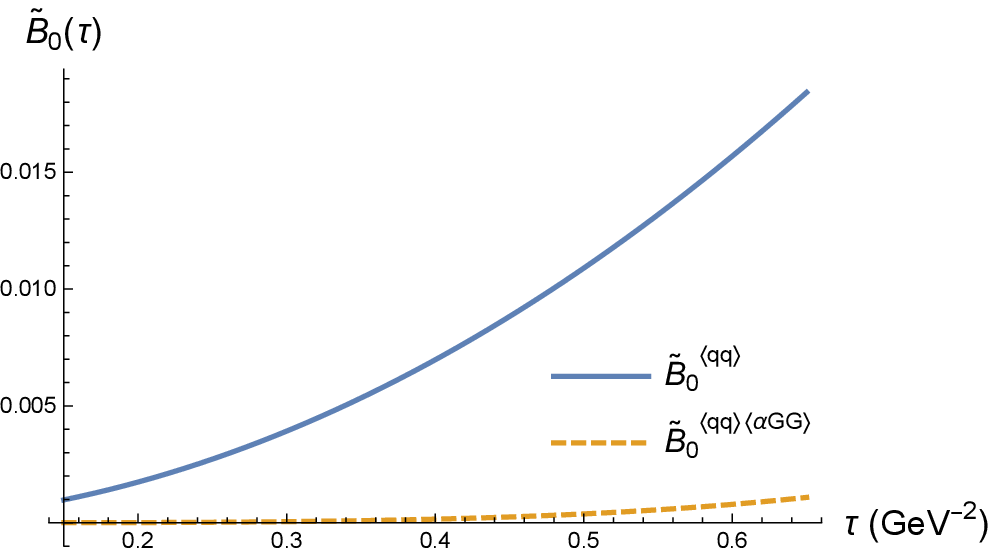}
    \end{minipage}
    \caption{\small{Ratios (left) $B_k$ of the gluon condensate $\langle \alpha_s GG\rangle$ to the LO (blue solid curves) and NLO (orange dashed curves) PT terms, and (right) $\tilde{B}_{k}$ of other non-PT terms to the NLO PT terms for $k=0$.}}
   \label{fig:B_k}
\end{figure}

In order to establish the Borel window, the criteria employed to get the upper bound on $\tau$ was based on the relative contribution of the QCD condensates to the Laplace sum-rule with respect to the PT terms
\vspace{-.2cm}
\begin{equation}
    B_k=\frac{{\cal L}^{\langle \alpha GG\rangle}_k(\tau)}{{\cal L}^{\rm PT}_k(\tau)}<\frac{1}{3}, \quad  \tilde B_k^{\langle \bar q q\rangle}=\frac{{\cal L}^{\langle \bar q q\rangle}_k(\tau)}{{\cal L}^{\rm PT}_k(\tau)},  \quad \tilde B_k^{\langle \bar q q\rangle\langle \alpha GG\rangle}=\frac{{\cal L}^{\langle \bar q q\rangle\langle \alpha GG\rangle}_k(\tau)}{{\cal L}^{\rm PT}_k(\tau)}.
    \label{B_k}
\end{equation}
The results from \eqref{B_k}, comparing the shift of the Borel window when NLO terms are added to the spectral function, are illustrated in Fig.~\ref{fig:B_k} and summarized in Table~\ref{borel_window_tab} (Left).
Here arises our first important conclusion: the inclusion of NLO PT terms in the spectral function is of great advantage for the QCDSR analysis, since it widens the Borel window by shifting the upper bound of $\tau$ (see Table~\ref{borel_window_tab}) enhancing the reliability of the analysis. 

\begin{table}[htb]
    \centering
    \begin{tabular}{l|r|r}
      $k$ & $B^{\langle \alpha GG\rangle/\rm{LO\, PT}}_{k}$ & $B^{\langle \alpha GG\rangle/\rm{NLO\, PT}}_{k}$\\\hline\hline
        0 &  $\tau\leq 0.47$ GeV$^{-2}$ & $\tau\leq 0.57$ GeV$^{-2}$ \\
        1 & $\tau\leq 0.61$ GeV$^{-2}$ & $\tau\leq 0.75$ GeV$^{-2}$ \\\hline
    \end{tabular}
    \qquad
    \begin{tabular}{l|c}
        Variable & Range\\\hline\hline
        $s_{0}$ & $\ge 0.33\,{\rm GeV^2}$\\
        $\tau$  &  $0.2\,\textrm{--} \,0.57\, {\rm GeV}^{-2}$
        \\
        \hline
    \end{tabular}
    \caption{(Left) Results of Eq.~\eqref{B_k} for $k=0,1$, showing the increase of the Borel window upper bound when adding the NLO PT corrections. (Right) Constraints on $s_0$ and $\tau$ defining the Borel-window working region for the QCDSR analysis.}
    \label{borel_window_tab}
\end{table}

Consecutively, setting a lower bound on $\tau$ was achieved by also using an anomalous dimension independent quantity, \textit{i.e.}, ratios of Laplace sum-rules. These corresponding conditions are standard integral inequalities, the so-called Schwarz \cite{Kleiv:2013dta} and H\"older inequalities \cite{Benmerrouche:1995qa}, respectively 
\vspace{-.2cm}
\begin{equation}
    \frac{\mathcal{R}_{k}(\tau,s_{0})/\mathcal{R}_{k-1}(\tau,s_{0})}{\mathcal{R}_{k-1}(\tau,s_{0})/\mathcal{R}_{k-2}(\tau,s_{0})}\ge 1,\, 
    \quad k\ge 2,
    \label{eq:holderIneq2}
\end{equation}
and
\vspace{-.2cm}
\begin{equation}
    \frac{\mathcal{R}_{k}\bigl(\tau+[1-\omega]\,\delta\tau,s_{0} \bigr)}{\mathcal{R}_{k}^{\omega}\bigl(\tau,s_{0}\bigr) \mathcal{R}_{k}^{1-\omega}\bigl(\tau+\delta\tau,s_{0}\bigr)}\leq 1,\,
    \quad 0\leq \omega\leq 1,
    \;\text{and } \;k\ge0\,.
    \label{eq:holderIneq1}
\end{equation}
Inequalities \eqref{eq:holderIneq1} and \eqref{eq:holderIneq2}, not only helped us compute a lower bound on the Borel parameter $\tau$, but also provided us the additional feature of constraining the continuum threshold $s_{0}$ from below.  
Our results for both parameters are shown in Table~\ref{borel_window_tab} (Right). Let us mention that the ratios of Laplace sum-rules in this stage showed a strong insensitivity to the presence of NLO contributions, as depicted in the unchanged lower bounds of $\tau$ and $s_{0}$.

\subsection{Resonance Models: Final Results} 
\begin{figure}[htb]
    \centering
    \includegraphics[scale=.8]{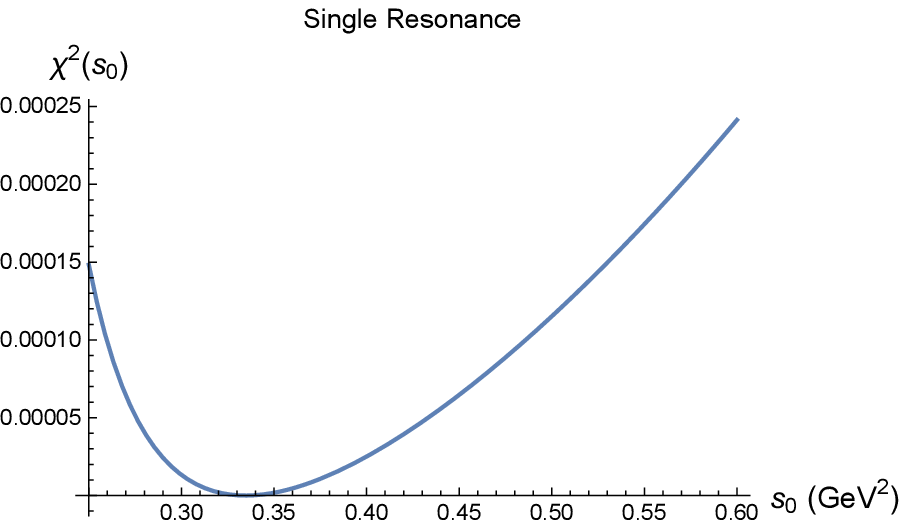}
    \includegraphics[scale=.8]{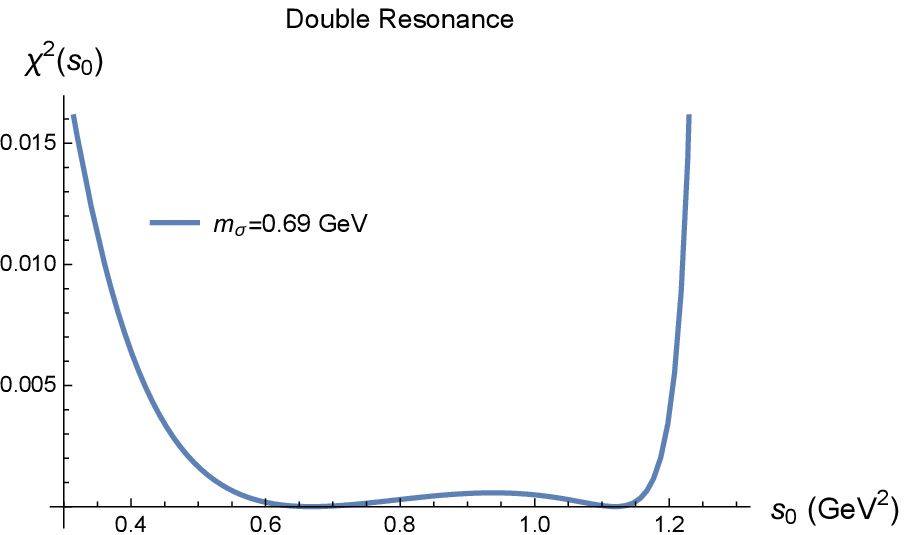}
    \caption{(Left) The SR model residual $\chi_{\text{SR}}^{2}$ shown as a function of $s_0$ for the optimized $m_\sigma$. (Right) The DR residual sum of squares \eqref{eq:chi2Double} shown as a function of $s_0$ for the optimized $m_\sigma=0.69\,{\rm GeV}$ (see Table~\ref{predictions_table}). }
    \label{dr_plots}
\end{figure}

Having obtained the proper Borel window to make a reliable QCDSR analysis, we can turn our attention to the models that will parameterize the coupling of resonance to the tetraquark current. 
Our first choice is the simple single-narrow resonance model (SR) outlined below, and their corresponding modification to ratios of \eqref{LSR_final}
\vspace{-.2cm}
\begin{equation}
    \rho^{\rm SR}(t)=A_\sigma\delta\left(t-m_\sigma^2\right),\quad\ \frac{\mathcal{R}_{1}(\tau,s_{0})}{\mathcal{R}_{0}(\tau,s_{0})} = m^{2}_{\sigma}\,.
    \label{eq:sr_model}
\end{equation}
The optimization method is based on the minimization of the residual sum of squares with respect to $m_{\sigma}$ and $s_{0}$,
\begin{equation}
        \chi_{\text{SR}}^{2}(s_{0})= \sum_{j} \Biggl( m_{\sigma}^{2} \frac{\mathcal{R}_{0}(\tau_{j},s_{0})}{\mathcal{R}_{1}(\tau_{j},s_{0})} -1\Biggr)^{2}\,.
    \label{eq:chi2Single}
\end{equation}
The optimized $m_\sigma$ can be expressed as a function of $s_0$, resulting in the single-variable residual shown in Fig.~\ref{dr_plots} (Left). Note that \eqref{eq:chi2Single} is defined by means of the Borel window, whose changes had also been explored, and the results of these variations are shown in Table~\ref{predictions_table}.

Our mass prediction for this state is in good agreement with the reported by \cite{ParticleDataGroup:2020ssz} for the $f_{0}(500)$ state. However, the optimized $s_{0}$ extracted from the analysis is located uncomfortably near the lower bound imposed by H\"{o}lder Inequalities (see Table~\ref{borel_window_tab}), leaving us with a reduced separation of the resonance location to the continuum. This shortcoming motivates the use of a double resonance model, where the extra heavier resonance included is the well-known $f_{0}(980)$. 
 
The double resonance model (DR) is then given by
\vspace{-.3cm}
\begin{equation}
\rho^{\rm DR}(t)=A_\sigma\delta\left(t-m_\sigma^2\right)+A_{f_0}\delta\left(t-m^2_{f_0}\right)\,,
\label{dr_model}    
\end{equation}
where $A_{f_0}>0$ parameterizes the coupling strength of the heavier state to the tetraquark current. Similarly, inserting \eqref{dr_model} into a ratio of \eqref{LSR_final}, we get
\vspace{-.5cm}
\begin{equation}
    \frac{\mathcal{R}_{1}(\tau,s_{0})}{\mathcal{R}_{0}(\tau,s_{0})} =  \frac{A_{\sigma}m_{\sigma}^{2} \,e^{-m_{\sigma}^{2}\tau} +A_{f_{0}}m_{f_{0}}^{2}\,e^{-m_{f_{0}}^{2}\tau}}{A_{\sigma} \,e^{-m_{\sigma}^{2}\tau} +A_{f_{0}}\,e^{-m_{f_{0}}^{2}\tau} }\,.
    \label{eq:QCDSRDouble}
\end{equation}

The mass of $f_{0}(980)$ was used as input in the optimization procedure. Given that $f_{0}(980)$ is a well-established state, and accounting the mass hierarchy with respect to $f_{0}(500)$ (or $\sigma$) as a compensation for the exponential suppression in \eqref{LSR_final}, the residual sum of squares becomes
\vspace{-.3cm}
\begin{gather}
    \chi_{\text{DR}}^{2}(s_{0})= \sum_{j} \Biggl( m_{\sigma}^{2}\frac{\mathcal{R}_{0}(\tau_{j},s_{0})}{\mathcal{R}_{1}(\tau_{j},s_{0})} - 1 + rm_{f_{0}}^{2}e^{-\Delta m^{2}\tau_{j}}\frac{\mathcal{R}_{0}(\tau_{j},s_{0})}{\mathcal{R}_{1}(\tau_{j},s_{0})} - re^{-\Delta m^{2}\tau_{j}} \Biggr)^{2}\,,
    \label{eq:chi2Double}
    \\
    r=\frac{A_{f_0}}{A_\sigma}\,,~\Delta m^2=m^2_{f_0}-m^2_\sigma\,.
\label{dr_parameters}
\end{gather}
The results from this optimization are shown in Table~\ref{predictions_table}, where $r$ was determined as a function of $s_{0}$ and $m_{\sigma}$. We obtained a mass prediction $m_{\sigma}=0.69\,\mathrm{GeV}$, again consistent with the range reported by \cite{ParticleDataGroup:2020ssz} for the $f_{0}(500)$ state, but now with a much better separation of the continuum threshold $s_{0}=1.12 \,\mathrm{GeV}^{2}$ from the heavier resonance $f_{0}(980)$. Additionally, we found similar patterns with Chiral Lagrangian analysis respect to the relative coupling strength \cite{Fariborz:2003uj, Fariborz:2009cq}, where the value of $r=3.38$ implies that the $f_{0}(980)$ coupling to the tetraquark current is favored over the $\sigma$ within the tetraquark picture. 

\begin{figure}[htb]
    \centering
    \includegraphics[scale=.8]{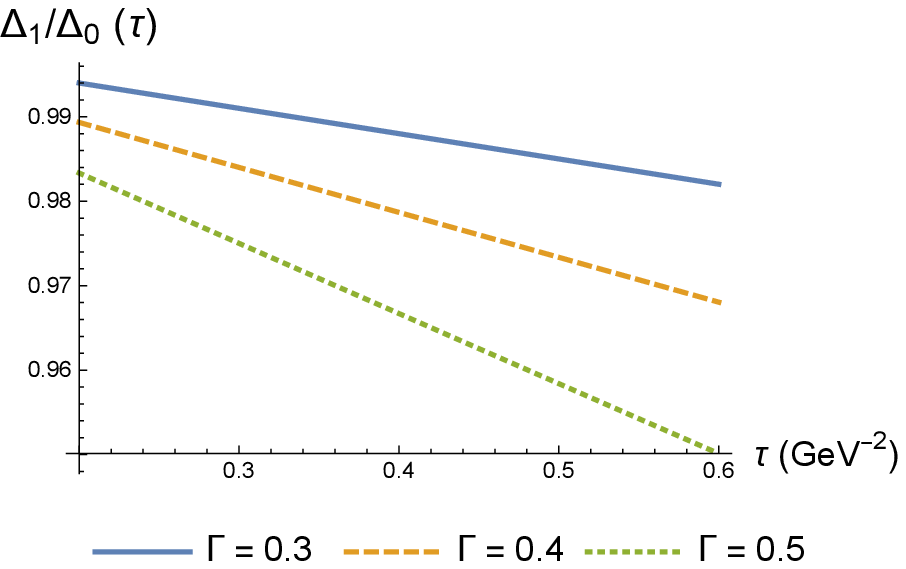}
    \includegraphics[scale=.8]{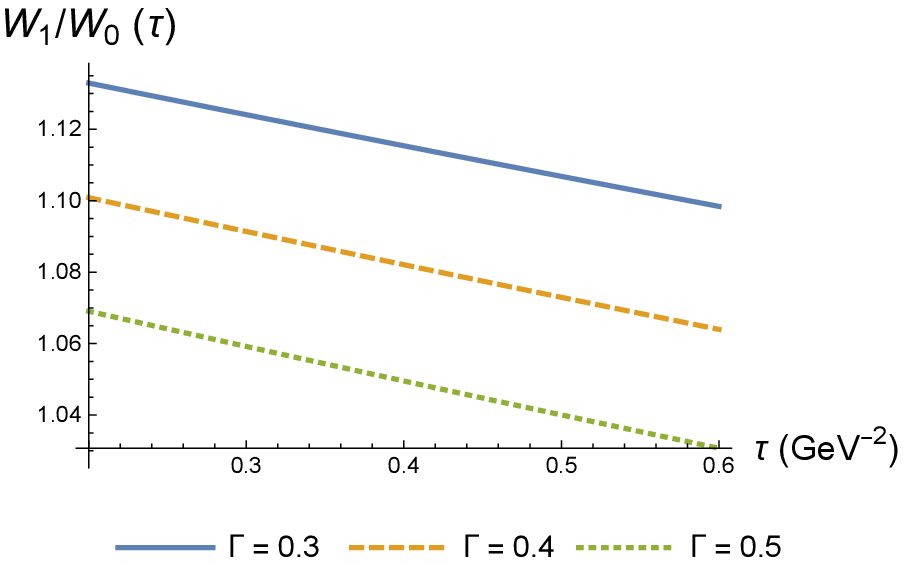}
    \caption{The quantity (left) $\Delta_{1}/\Delta_{0}$, and (right) $W_1/W_0$ is shown as a function of $\tau$ within the Borel window for $m_\sigma=0.69\,{\rm GeV}$ and for selected values of $\Gamma_\sigma$.}
    \label{fig:Assym_shape_DR}
\end{figure}

\begin{table}[htb]
\centering
    \begin{tabular}{l|ccrcc}
    Model  & Range $s_{0}$ (GeV$^{2}$)  & Range $\tau$ (GeV$^{-2}$) & $m_{\sigma}$ (GeV) & $s_0^{\text{opt}}$ (GeV$^{2}$) & $r$  \\\hline\hline
    SR & $0.33$ 
    -- $1.3$ 
    & $0.2 - 0.57$~(${\pm 0.05}$) & $0.52\pm0.01$ & 0.335 &-- \\
    DR & $0.98$ -- $1.3$ & $0.2$ -- $0.57$~($\pm 0.05$) & $0.69\pm0.03$  & 1.12  & $3.38$ \\\hline
\end{tabular}
\caption{Predictions for the optimized mass and continuum threshold in the SR and DR models. With the inclusion of resonance width effects, $m_\sigma$ and $r$ Table entries are interpreted as the effective mass and coupling ratio $m^{\rm eff}_\sigma$ and $r^{\rm eff}$. }
\label{predictions_table}
\end{table}
Furthermore, as $f_{0}(500)$ is considered a broad state \cite{ParticleDataGroup:2020ssz}, our analysis contemplated models including symmetric and asymmetric width ($\Gamma_{\sigma}$) shapes. The former, symmetric width model (SW) parameterizes the spectral function as 
\begin{equation}
\rho^{\rm SW}(t)=\frac{A_{\sigma}}{2m_{\sigma} \Gamma_{\sigma}}
\left[
\theta\left(t-m_\sigma^2+m_{\sigma}\Gamma_{\sigma}\right) 
-\theta\left(t-m_\sigma^2-m_{\sigma}\Gamma_{\sigma}\right)
\right]\,.
\label{square_pulse_rho}
\end{equation}
The resulting SW modification \cite{Elias:1998bq} of the SR model \eqref{eq:sr_model}, and to the DR model \eqref{eq:QCDSRDouble} are given by
\begin{equation}
    \mathrm{SR} \rightarrow \quad\frac{\mathcal{R}_{1}(\tau,s_{0})}{\mathcal{R}_{0}(\tau,s_{0})} = m^{2}_{\sigma}\frac{\Delta_1\left(m_{\sigma},\Gamma,\tau\right)}{\Delta_0\left(m_{\sigma},\Gamma,\tau\right)}
    \,,
    \label{SW_single}
\end{equation}
\begin{equation}
    \mathrm{DR}\rightarrow\quad \frac{\mathcal{R}_{1}(\tau,s_{0})}{\mathcal{R}_{0}(\tau,s_{0})} =
    \frac{\Delta_0 A_{\sigma}\left(m_{\sigma}^{2}\frac{\Delta_1}{\Delta_0}\right) \,e^{-m_{\sigma}^{2}\tau} +A_{f_{0}}m_{f_{0}}^{2}\,e^{-m_{f_{0}}^{2}\tau}}{\Delta_0 A_{\sigma} \,e^{-m_{\sigma}^{2}\tau} +A_{f_{0}}\,e^{-m_{f_{0}}^{2}\tau} },
    \label{SW_double}    
\end{equation}
where
\vspace{-.3cm}
\begin{gather}
    \Delta_{0}\left(m_{\sigma},\Gamma,\tau\right) = \frac{\sinh{\left(m_{\sigma}\Gamma_{\sigma}\tau\right)}}{m_{\sigma}\Gamma_{\sigma}\tau},
    \\[5pt]
    \Delta_{1}\left(m_{\sigma},\Gamma,\tau\right) = \left(1+ \frac{1}{m_{\sigma}^{2}\tau}\right)
    \Delta_{0}\left(m_{\sigma},\Gamma,\tau\right)- \frac{\cosh{\left(m_{\sigma}\Gamma_{\sigma}\tau\right)}}{m_{\sigma}^{2}\tau}\,.    
\end{gather}
It was found that for values of $0.2\,\mathrm{GeV}<\Gamma_{\sigma}<0.6\,\mathrm{GeV}$, and the parameters given in Table~\ref{predictions_table}, the ratio $\Delta_{1}/\Delta_{0}$ is approximately $1$ in both models (SR and DR), as shown in Fig.~\ref{fig:Assym_shape_DR} (Left), thus leaving the door open to consider the asymmetric width shape model. 

Later, for an asymmetric width shape (AW) that could be obtained from Chiral Lagrangian methods \cite{Sannino:1995ik, Harada:1995dc, Black:1999dx}, the resonance models are modified by
\vspace{-.2cm}
\begin{equation}
   \rho^{\rm AW}(s) = \frac{A_\sigma}{N m^2_{\sigma} }\frac{s^{2}}{\left[(s-m_{\sigma}^{2})^{2} + \frac{\Gamma^{2}_{\sigma}}{m_{\sigma}^{6}}s^{4}\right]},
   \label{rho_AW}
\end{equation}
where $N$ is a normalization factor to ensure that \eqref{rho_AW} and \eqref{eq:sr_model} have the same integrated resonance strength. The AW modifications to SR and DR models, analogous to \eqref{SW_single} and \eqref{SW_double} are 
\begin{equation}
    \mathrm{SR}\rightarrow \quad\frac{\mathcal{R}_{1}(\tau,s_{0})}{\mathcal{R}_{0}(\tau,s_{0})} = m^{2}_{\sigma}\frac{W_1\left(m_{\sigma},\Gamma,\tau,s_0\right)}{W_0\left(m_{\sigma},\Gamma,\tau,s_0\right)}
    \;,
    \label{AW_single}
\end{equation}
\begin{equation}
    \mathrm{DR} \rightarrow\quad \frac{\mathcal{R}_{1}(\tau,s_{0})}{\mathcal{R}_{0}(\tau,s_{0})} =
    \frac{W_0 A_{\sigma}\left(m_{\sigma}^{2}\frac{W_1}{W_0}\right) \,e^{-m_{\sigma}^{2}\tau} +A_{f_{0}}m_{f_{0}}^{2}\,e^{-m_{f_{0}}^{2}\tau}}{W_0 A_{\sigma} \,e^{-m_{\sigma}^{2}\tau} +A_{f_{0}}\,e^{-m_{f_{0}}^{2}\tau} },
    \label{AW_double}
\end{equation}
where 
\vspace{-.3cm}
\begin{equation}
    W_{k} \left(m_{\sigma},\Gamma_{\sigma},\tau,s_{0}\right) =\frac{1}{N} \int_{-1}^{\frac{s_{0}-m_{\sigma}^{2}}{m_{\sigma}^{2}}} \,d\zeta\,e^{-m_{\sigma}^{2}\zeta\tau}\,\frac{(\zeta+1)^{2+k}}{\zeta^{2}+\frac{\Gamma_{\sigma}^{2}}{m_{\sigma}^{2}}(\zeta+1)^{4}}\,,\quad \mathrm{for }\;\, k=0,1.
\end{equation}
The results of the effects of the asymmetric width as a function of the Borel window are showed in Fig.~\ref{fig:Assym_shape_DR} (Right). They resulted in a suppression on effective mass of around $5\%$ compared to the SW, being consistent with the trend found in \cite{Shi:1999hm}. Though the ratio $W_{1}/W_{0}$ as a function of $\tau$ seems to be non-negligible, the effective mass term appears to be surprisingly robust under the width effects. 

\section{Conclusions}
In this paper, we performed a comprehensive QCDSR analysis of the lightest tetraquark scalar $f_{0}(500)/\sigma$. The study focused on the impact of NLO perturbative terms compared to the LO on the mass prediction, using specifically ratios of Laplace sum-rules. The inclusion of these NLO corrections in the hadronic spectral function was crucial, as it benefited the analysis by shifting the upper bound on the Borel window, making the QCDSR analysis more reliable. 

NLO corrections demonstrated large contributions to the individual Laplace sum-rules, but these were compensated when using ratios of them. Nevertheless, the great impact of these NLO terms were somewhat overshadowed with the use of the ratios of Laplace sum-rules, they had the additional feature of removing dependence on the anomalous dimension. 

As a summary, the NLO tetraquark QCD Laplace sum-rules were employed in a variety of models to obtain a mass prediction of the $\sigma$ state, which included single and double resonances, as well as symmetric and asymmetric width shapes. We concluded that the mass prediction depends indirectly on these corrections, via the widened Borel window shown as in Table~\ref{borel_window_tab}. We also concluded that the single narrow resonance model was not enough to understand the $\sigma$ state, given that the continuum threshold was too close to the permitted by H\"{o}lder Inequalities, motivating further studies with more complex models. Additionally, we noticed the mass prediction was surprisingly robust under the inclusion of widths. Finally, we were then able to obtain results that are in good agreement with mass values given in \cite{ParticleDataGroup:2020ssz}, and from the double resonances model, we computed a coupling strength consistent with the findings from Chiral Lagrangian studies \cite{Fariborz:2003uj, Fariborz:2009cq}. See Ref.~\cite{Cid-Mora:2022kgu} for additional detail of our work.


\begin{thebibliography}{99}
%\bibliography{Biblio} 
%\bibliographystyle{h-physrev}
%\cite{Cid-Mora:2022kgu}
\bibitem{Cid-Mora:2022kgu}
B.~A.~Cid-Mora and T.~G.~Steele,
%``Next-to-leading order (NLO) perturbative effects in QCD sum-rule analyses of light tetraquark systems: A case study in the scalar-isoscalar channel,''
Nucl. Phys. A \textbf{1028}, 122538 (2022),
%%doi:10.1016/j.nuclphysa.2022.122538
2206.06280.
%1 citations counted in INSPIRE as of 16 Jan 2023

%\cite{ParticleDataGroup:2020ssz}
\bibitem{ParticleDataGroup:2020ssz}
Particle Data Group, P.~A.~Zyla \textit{et al.}, 
%``Review of Particle Physics,''
PTEP \textbf{2020}, 083C01 (2020).
%%doi:10.1093/ptep/ptaa104
%4919 citations counted in INSPIRE as of 16 Jan 2023

%\cite{Belle:2003nnu}
\bibitem{Belle:2003nnu}
Belle, S.~K.~Choi \textit{et al.},
%``Observation of a narrow charmonium-like state in exclusive $B^\pm \to K^\pm \pi^+ \pi^- J/\psi$ decays,''
Phys. Rev. Lett. \textbf{91}, 262001 (2003), 
%%doi:10.1103/PhysRevLett.91.262001
hep-ex/0309032.
%2250 citations counted in INSPIRE as of 16 Jan 2023

%\cite{Belle:2007hrb}
\bibitem{Belle:2007hrb}
Belle, S.~K.~Choi \textit{et al.},
%``Observation of a resonance-like structure in the $pi^\pm \psi^\prime$ mass distribution in exclusive $B \to K \pi^\pm \psi^\prime$ decays,''
Phys. Rev. Lett. \textbf{100}, 142001 (2008), 
%doi:10.1103/PhysRevLett.100.142001
0708.1790.
%763 citations counted in INSPIRE as of 16 Jan 2023

%\cite{BESIII:2013ris}
\bibitem{BESIII:2013ris}
BESIII, M.~Ablikim \textit{et al.},
%``Observation of a Charged Charmoniumlike Structure in $e^+e^- \to \pi^+\pi^- J/\psi$ at $\sqrt{s}$ =4.26 GeV,''
Phys. Rev. Lett. \textbf{110}, 252001 (2013), 
%doi:10.1103/PhysRevLett.110.252001
1303.5949.
%998 citations counted in INSPIRE as of 16 Jan 2023

%\cite{Belle:2013yex}
\bibitem{Belle:2013yex}
Belle, Z.~Q.~Liu \textit{et al.},
%``Study of $e^+e^- → π^+ π^- J/ψ$ and Observation of a Charged Charmoniumlike State at Belle,''
Phys. Rev. Lett. \textbf{110}, 252002 (2013), 
[erratum: Phys. Rev. Lett. \textbf{111}, 019901 (2013)], 
%doi:10.1103/PhysRevLett.110.252002
1304.0121.
%820 citations counted in INSPIRE as of 16 Jan 2023

%\cite{LHCb:2021auc}
\bibitem{LHCb:2021auc}
LHCb, R.~Aaij \textit{et al.},
%``Study of the doubly charmed tetraquark $T_{cc}^{+}$,''
Nature Commun. \textbf{13}, 3351 (2022), 
%doi:10.1038/s41467-022-30206-w
2109.01056.
%152 citations counted in INSPIRE as of 16 Jan 2023

%\cite{LHCb:2021vvq}
\bibitem{LHCb:2021vvq}
LHCb, R.~Aaij \textit{et al.},
%``Observation of an exotic narrow doubly charmed tetraquark,''
Nature Phys. \textbf{18}, 751-754 (2022), 
%doi:10.1038/s41567-022-01614-y
2109.01038.
%170 citations counted in INSPIRE as of 16 Jan 2023

%\cite{Jaffe:1976ig}
\bibitem{Jaffe:1976ig}
R.~L.~Jaffe,
%``Multi-Quark Hadrons. 1. The Phenomenology of (2 Quark 2 anti-Quark) Mesons,''
Phys. Rev. D \textbf{15}, 267 (1977).
%doi:10.1103/PhysRevD.15.267
%2150 citations counted in INSPIRE as of 16 Jan 2023

%\cite{Shifman:1978bx}
\bibitem{Shifman:1978bx}
M.~A.~Shifman, A.~I.~Vainshtein and V.~I.~Zakharov,
%``QCD and Resonance Physics. Theoretical Foundations,''
Nucl. Phys. B \textbf{147}, 385-447 (1979).
%doi:10.1016/0550-3213(79)90022-1
%5626 citations counted in INSPIRE as of 16 Jan 2023

%\cite{Shifman:1978by}
\bibitem{Shifman:1978by}
M.~A.~Shifman, A.~I.~Vainshtein and V.~I.~Zakharov,
%``QCD and Resonance Physics: Applications,''
Nucl. Phys. B \textbf{147}, 448-518 (1979).
%doi:10.1016/0550-3213(79)90023-3
%3083 citations counted in INSPIRE as of 16 Jan 2023

%\cite{Groote:2014pva}
\bibitem{Groote:2014pva}
S.~Groote, J.~G.~K\"orner and D.~Niinepuu,
%``Perturbative $O(\alpha_s)$ corrections to the correlation functions of light tetraquark currents,''
Phys. Rev. D \textbf{90}, 054028 (2014), 
%doi:10.1103/PhysRevD.90.054028
1401.4801.
%6 citations counted in INSPIRE as of 16 Jan 2023

%\cite{Chen:2007xr}
\bibitem{Chen:2007xr}
H.~X.~Chen, A.~Hosaka and S.~L.~Zhu,
%``Light Scalar Tetraquark Mesons in the QCD Sum Rule,''
Phys. Rev. D \textbf{76}, 094025 (2007), 
%doi:10.1103/PhysRevD.76.094025
0707.4586.
%97 citations counted in INSPIRE as of 16 Jan 2023

%\cite{Narison:1981ts}
\bibitem{Narison:1981ts}
S.~Narison and E.~de Rafael,
%``On QCD Sum Rules of the Laplace Transform Type and Light Quark Masses,''
Phys. Lett. B \textbf{103}, 57-62 (1981).
%doi:10.1016/0370-2693(81)90193-3
%193 citations counted in INSPIRE as of 16 Jan 2023

%\cite{Reinders:1984sr}
\bibitem{Reinders:1984sr}
L.~J.~Reinders, H.~Rubinstein and S.~Yazaki,
%``Hadron Properties from QCD Sum Rules,''
Phys. Rept. \textbf{127}, 1 (1985).
%doi:10.1016/0370-1573(85)90065-1
%1802 citations counted in INSPIRE as of 16 Jan 2023

%\cite{Narison:2002woh}
\bibitem{Narison:2002woh}
S.~Narison,
%``QCD as a Theory of Hadrons,''
Camb. Monogr. Part. Phys. Nucl. Phys. Cosmol. \textbf{17}, 1-812 (2007)
Cambridge University Press, 2022, 
%ISBN 978-1-00-929029-6, 978-1-00-929031-9, 978-1-00-929033-3, 978-0-521-03731-0, 978-0-521-81164-4, 978-0-511-18948-7
%doi:10.1017/9781009290296
hep-ph/0205006.
%290 citations counted in INSPIRE as of 16 Jan 2023

%\cite{Gubler:2018ctz}
\bibitem{Gubler:2018ctz}
P.~Gubler and D.~Satow,
%``Recent Progress in QCD Condensate Evaluations and Sum Rules,''
Prog. Part. Nucl. Phys. \textbf{106}, 1-67 (2019), 
%doi:10.1016/j.ppnp.2019.02.005
1812.00385.
%63 citations counted in INSPIRE as of 16 Jan 2023

%\cite{Kleiv:2013dta}
\bibitem{Kleiv:2013dta}
R.~T.~Kleiv, T.~G.~Steele, A.~Zhang and I.~Blokland,
%``Heavy-light diquark masses from QCD sum rules and constituent diquark models of tetraquarks,''
Phys. Rev. D \textbf{87}, 125018 (2013), 
%doi:10.1103/PhysRevD.87.125018
1304.7816.
%75 citations counted in INSPIRE as of 16 Jan 2023

%\cite{Benmerrouche:1995qa}
\bibitem{Benmerrouche:1995qa}
M.~Benmerrouche, G.~Orlandini and T.~G.~Steele,
%``Constraints on QCD sum rules from the Holder inequalities,''
Phys. Lett. B \textbf{356}, 573-579 (1995), 
%doi:10.1016/0370-2693(95)00875-L
hep-ph/9507304.
%19 citations counted in INSPIRE as of 16 Jan 2023

%\cite{Fariborz:2003uj}
\bibitem{Fariborz:2003uj}
A.~H.~Fariborz,
%``Isosinglet scalar mesons below 2-GeV and the scalar glueball mass,''
Int. J. Mod. Phys. A \textbf{19}, 2095-2112 (2004), 
%doi:10.1142/S0217751X04018695
hep-ph/0302133.
%103 citations counted in INSPIRE as of 16 Jan 2023

%\cite{Fariborz:2009cq}
\bibitem{Fariborz:2009cq}
A.~H.~Fariborz, R.~Jora and J.~Schechter,
%``Global aspects of the scalar meson puzzle,''
Phys. Rev. D \textbf{79}, 074014 (2009), 
%doi:10.1103/PhysRevD.79.074014
0902.2825.
%99 citations counted in INSPIRE as of 16 Jan 2023

%\cite{Elias:1998bq}
\bibitem{Elias:1998bq}
V.~Elias, A.~H.~Fariborz, F.~Shi and T.~G.~Steele,
%``QCD sum rule consistency of lowest lying q anti-q scalar resonances,''
Nucl. Phys. A \textbf{633}, 279-311 (1998), 
%doi:10.1016/S0375-9474(98)00119-5
hep-ph/9801415.
%81 citations counted in INSPIRE as of 16 Jan 2023

%\cite{Sannino:1995ik}
\bibitem{Sannino:1995ik}
F.~Sannino and J.~Schechter,
%``Exploring pi pi scattering in the 1/N(c) picture,''
Phys. Rev. D \textbf{52}, 96-107 (1995), 
%doi:10.1103/PhysRevD.52.96
hep-ph/9501417.
%110 citations counted in INSPIRE as of 16 Jan 2023

%\cite{Harada:1995dc}
\bibitem{Harada:1995dc}
M.~Harada, F.~Sannino and J.~Schechter,
%``Simple description of pi pi scattering to 1-GeV,''
Phys. Rev. D \textbf{54}, 1991-2004 (1996), 
%doi:10.1103/PhysRevD.54.1991
hep-ph/9511335.
%245 citations counted in INSPIRE as of 16 Jan 2023

%\cite{Black:1999dx}
\bibitem{Black:1999dx}
D.~Black, A.~H.~Fariborz and J.~Schechter,
%``Chiral Lagrangian treatment of pi eta scattering,''
Phys. Rev. D \textbf{61}, 074030 (2000), 
%doi:10.1103/PhysRevD.61.074030
hep-ph/9910351.
%59 citations counted in INSPIRE as of 16 Jan 2023
 %\cite{Shi:1999hm}
\bibitem{Shi:1999hm}
F.~Shi, T.~G.~Steele, V.~Elias, K.~B.~Sprague, Y.~Xue and A.~H.~Fariborz,
%``Holder inequalities and isospin splitting of the quark scalar mesons,''
Nucl. Phys. A \textbf{671}, 416-446 (2000), 
%doi:10.1016/S0375-9474(99)00837-4
hep-ph/9909475.
%37 citations counted in INSPIRE as of 16 Jan 2023


\end{thebibliography}
\end{document}